\documentclass[aps,pre,twocolumn,showpacs]{revtex4}
\usepackage{epsfig}
\usepackage{graphicx}
\usepackage{bm}
\usepackage{amssymb}
\usepackage{amsmath}
\usepackage{amsthm}
\usepackage{color}
\begin{document}
\definecolor{brown}{rgb}{0.74,0.56,0.56}
\definecolor{purple}{rgb}{0.58,0,0.82}
\title{Correlations in Bipartite Collaboration Networks}
\author{Matti Peltom\"aki and Mikko Alava}
\affiliation{
Laboratory of Physics, Helsinki University of Technology, P.O.Box
1100, 02015 HUT, Finland}
\pacs{89.75.Hc, 87.23.Ge, 05.70.Ln}
\date{\today}

\begin{abstract}
Collaboration networks are studied as an example of growing
bipartite networks. These have been previously observed to exhibit
structure such as positive correlations between nearest-neighbour degrees.
However, a detailed understanding of the origin of such and 
the growth dynamics is lacking. Both of these issues are analyzed empirically
and simulated using various models. A new growth model is presented,
incorporating empirically necessary ingredients such as 
bipartiteness and sublinear preferential attachment.
This, and a recently proposed model of team assembly both
agree roughly with some empirical observations and fail
in several others.
\end{abstract}

\maketitle
\section{Introduction}

The study of networks has gained much attention in the physics literature
recently \cite{dorogovtsev:book, dorogovtsev:advphys51, newman:siamreview45, 
albert:revmodphys74}. The physics view on networks is to consider them
using the tools of  statistical mechanics.
The availability of large databases 
has made it possible to do empirical studies of large
networks of different disciplines. A number of such networks
have been identified and analyzed in the literature, the emphasis
being mostly on the basic characteristics of the networks, such as
the degree distribution, the clustering coefficient and the average
shortest path length. However, it has also been observed that
the degrees of nearest neighbour nodes are 
not statistically independent but mutually correlated
in practically every network imaginable
\cite{pastor-satorras:prl87, vazguez:pre65, newman:prl89, newman:pre67}.
In empirical observations, it is typically found that technological and
biological networks have negative correlations, also termed 
dissortative mixing, whereas social networks tend to have positive
correlations \cite{newman:pre68}. The forming of triads 
(i.e.~fully connected triplets) \cite{holme:pre65}, network bipartiteness
\cite{guillaume:cond-mat0307} and a hierarchical structure of social
networks \cite{boguna:cond-mat0309} have been suggested as reasons for
the assortative mixing. It has also been found out that the presence of 
correlations might have consequences regarding the physics of
dynamical models on networks \cite{boguna:pre66, boguna:prl90}.

In this article, we take a close, empirical look at the degree-degree correlation
structure of social collaboration networks. These networks are by force
bipartite, in contrast to many others.  
A bipartite graph is a graph with two kinds of vertices, say, $A$ and 
$T$, in which there are only edges between two vertices 
of different kinds. The $A$ nodes can be thought 
of as social actors or collaborators and the $T$ nodes as social ties or
collaboration acts. Typical examples of these networks are the movie-actor
network (the movies are the collaboration acts) and scientist-article 
networks where the scientists (the collaborators) appear together as
authors on the articles which play the role of collaboration acts. 

From a bipartite network, one can construct its unipartite counterpart,
the so-called one-mode projection onto actors (ties), as a network consisting
solely of the actors (ties) as nodes, two of which are connected by an edge
for each social tie (actor) they both participate in (enlist as participants).
For example, in the one-mode projection two scientists
are connected to each other as many times as they have co-authored a 
paper (an alternative definition not considered here would be to 
use this to define a weight for the link).

Three important questions arise in this context. First, what is
the structure of the bipartite network? The relevant quantities
are stated in the next section. Second, what can be stated in
general of the one-mode projection graph and its correlations?
We consider this mostly via the average nearest-neighbour degree (ANND).
Here, there is the main empirical observation that the ANND follows
a power-law scaling, when considered as a function of the degree of 
the central node. Moreover the degree distributions decay faster 
than scale-free ones and we discover sublinear effective preferential 
attachment  (PA) rules, independent of time.

Third, we consider two models. First,
a growing bipartite network model is introduced such that it
incorporates sublinear preferential attachment.
We perform simulations of this model, and a team assembly model introduced
recently by Guimer\'a \emph{et al.} \cite{guimera:science308}.
Both models can reproduce roughly the one-mode actor degree distributions
and the latter also the power-law scaling of the actor ANND. However, the
assembly model fails in matching the sublinear (empirical) PA rule, and
in matching the clustering as such.

This paper is organized as follows. Section 2 discusses the quantities measuring
network topology. In Section 3,
results of empirical measurements are presented. Section 4 visits
earlier models with similar goals. A new one is introduced in Section 5. 
In Section 6, the new model and the earlier ones are compared to empirical
measurements. 
Finally, Section 7 ends the paper with discussion and conclusions.

\section{Network topology}

Let $P(k)$ be the degree distribution in the one-mode projection onto actors,
i.e.~the probability that a randomly selected actor has $k$ links. 
This quantity often exhibits a fat tail that can be approximated with a
power-law. The degree-degree correlations in the networks are seen from
the joint probability distribution $P(k,k')$ where 
$(2-\delta_{k,k'})P(k,k')$ is the probability that a randomly selected edge
connects nodes with degree $k$ and $k'$. In undirected graphs (which are
considered here), $P(k,k')$ is necessarily symmetric with respect to $k$
and $k'$. In uncorrelated networks, it takes the form
\begin{equation} \label{eq:joint_uncorrelated}
P(k,k') = \frac{kk'P(k)P(k')}{\langle k \rangle^2} \, .
\end{equation}

The joint distribution $P(k,k')$ is often hard to measure empirically
due to a lack of a representative sample, i.e.~in real-life networks
there are typically only a few edges connecting nodes with given degrees 
$k$ and $k'$. Thus, another measures for the correlations have been 
devised, the most important of these being the average nearest-neighbour
degree (ANND), which is the average degree of the nearest neighbours
of nodes of degree $k$. Subsequently, it can be expressed as 
\begin{equation} \label{eq:annd_def}
\overline{k}_{nn} (k) = \langle k \rangle 
\frac{\sum_{k'} k' P(k,k')}{kP(k)} \, .
\end{equation}
This quantity is less vulnerable to statistical fluctuations than $P(k,k')$
but naturally less informative.

The degree-degree correlations can also be described by a Pearson
correlation coefficient $r$ between nearest-neighbour degrees. 
It is defined as
\cite{newman:prl89, ramasco:pre70}
\begin{equation} \label{eq:pearson_correlation_coefficient}
r = \langle k \rangle 
\frac{\sum_k k^2 \overline{k}_{nn}(k) P(k) - \langle k^2\rangle^2}
{\langle k \rangle \langle k^3 \rangle - \langle k^2\rangle^2} \, .
\end{equation}
If the network is uncorrelated, the ANND is a constant
$\overline{k}_{nn}(k)= \langle k^2\rangle / \langle k \rangle$ and 
the correlation coefficient vanishes. A positive value of $r$ and 
an increasing $\overline{k}_{nn}(k)$ are signs of assortative mixing. 

Another important quantity in networks is clustering or network transitivity,
which is a measure of the tendency to find fully connected triangles in the 
graph. It can be measured from
several different perspectives and the terminology
in the literature varies between different sources. 
Here, notation and terminology adapted
from Ref.~\cite{dorogovtsev:pre69} is used and is as follows. 

Let $m_{nn}(x)$ be the number of links between the nearest neighbours of 
a given vertex $x$ with degree $k$. The maximum number of such 
connections is $k(k-1)/2$. Define the local clustering of node $x$ as
\begin{equation} \label{eq:local_clustering}
C_x = \frac{m_{nn}(x)}{k(k-1)/2} \, .
\end{equation}

Now, the global clustering characteristics of the network are the following.
\begin{itemize}
\item The degree-dependent clustering. This is the average of the local clustering
of nodes with a given degree, i.e.
\begin{equation} \label{eq:k_clustering}
C(k) = \langle C_x \rangle_k \, ,
\end{equation}
where the subscript $k$ emphasizes the fact that the average is taken only over
nodes $x$ with degree $k$. 
\item The average clustering, which is the average of the local clustering over
all nodes in the graph. It is defined in terms of the degree distribution $P(k)$
and the degree-dependent clustering $C(k)$ as
\begin{equation} \label{eq:average_clustering}
\overline{C} = \sum_k P(k)C(k) \, .
\end{equation}
\item The clustering coefficient which is three times the ratio of the total
number of loops of length three in the graph to the total number of connected
triplets of vertices. It can also be defined in terms of $P(k)$ and $C(k)$ as
\begin{equation} \label{eq:clustering_coefficient}
c = \frac{\sum_k k(k-1)P(k)C(k)}{\langle k^2 \rangle - \langle k\rangle} \, .
\end{equation}
\end{itemize}

In networks without degree-degree correlations, the three clustering
characteristics in Eqs.~(\ref{eq:k_clustering}), (\ref{eq:average_clustering})
and (\ref{eq:clustering_coefficient}) equal each other 
\cite{dorogovtsev:pre69}
\begin{equation}
C(k) = \overline{C} = c = 
\frac{(\langle k^2 \rangle - \langle k \rangle)^2}
{N\langle k \rangle^3} \, ,
\end{equation}
where $N$ is the number of vertices in the network.

In this work, the emphasis is on the ANND and the $k$-dependent clustering. 
These metrics probe degree-degree correlations and the density of closed
loops of length three, respectively, which are considered important local
characteristics of networks. These properties could also be measured by using
the assortativity coefficient $r$ and the clustering coefficient $c$. These
differ, however, from those chosen to be emphasized here in an important 
respect; ANND and $C(k)$ provide more detailed information on the network
structure than $r$ and $c$, which are merely scalar quantities that can 
assume same values for several different correlation or clustering profiles. 
Furthermore, the statistical quality of the empirical networks appears to be
high enough for these quantities to be reliably measured. 

\section{Empirical results}

\subsection{Analyzed networks}

In this work, the empirically analyzed data comes from two sources:
from the Internet Movie Database (IMDB) \cite{imdb} 
(an older but preprocessed data set is also available at the web site 
\cite{yeong:imdb-data})
and from the arXiv.org preprint server \cite{arxiv}.

The actor--movie network from the IMDB is a bipartite network consisting
of actors and movies (social ties) where an actor is linked to a movie
if he acted in it. The network is rather comprehensive,
containing around 770 000 actors in about 430 000 films, the oldest
one of which dates
back to 1890. The IMDB reports the on-screen credits as its primary
data source. 

The arXiv.org preprint server hosts a collection of electronically
available preprints in several disciplines of physics and related
sciences. From such data, a bipartite graph of scientists (social actors) and
articles (social ties) can be constructed. It is 
reasonable to assume that different disciplines are rather 
disconnected when author collaboration is considered. Thus it is natural to 
analyze them separately. For this, three different
disciplines, which contain most of the articles stored in the database,
were chosen, namely astrophysics (astro-ph), condensed matter physics
(cond-mat) and the phenomenology of high energy physics (hep-ph). The 
number of articles in other disciplines is not large enough to permit
a meaningful data analysis. The networks analyzed here contain articles up to
the end of 2003. Note that though in the bipartite graph each edge
is unique, multiple ties shared between the same pair of actors will
produce multiple, degenerate links.

Denote the degrees of actors and ties in the full bipartite representation
by $q_a$ and $q_t$, respectively, and their one-mode projected counterparts
by $k$ (for the actors) and $k_t$ (for the ties). 
The basic parameters of the four empirical networks 
under study
can be found in Table 
\ref{table:parameters}. 
The values of the clustering coefficient, the
average clustering and the assortativity
coefficient differ from those in 
Refs.~\cite{ramasco:pre70,newman:prl89,barabasi:physicaa311}, since
newer versions of the data are used. The connectivity of the network
was also studied, leading to the conclusion that all four networks
consist of a giant component and a very small number of nodes outside it;
the second largest component in the 
condensed matter network is composed of 19 scientists, for instance. 
In other words, they are far from any kind of percolation transition
whether in the bipartite form or in the one-mode projection.

\begin{table}[t]
\begin{tabular}{|l|r|r|r|r|}
\hline
network & actor & astro-ph & cond-mat & hep-ph \\
\hline \hline
$N_a$ & 766 386 & 21 843 & 28 526 & 11 343 \\
$N_t$ & 427 969 & 47 580 & 49 330 & 39 382 \\
$\langle q_a \rangle$ & 3.68 & 10.4 & 5.08 & 7.89 \\
$\langle q_t \rangle$ & 6.59 & 4.77 & 2.94 & 2.28 \\
$\langle k \rangle$ & 87.4 & 57.2 & 16.0 & 23.4 \\
$\langle k_t \rangle$ & 137.8 & 118.3 & 56.3 & 70.4 \\
$r$ & 0.292 & 0.433 & 0.250 & 0.344 \\
$c$ & 0.27 & 0.578 & 0.370 & 0.441 \\
$\overline{C}$ & 0.817 & 0.683 & 0.674 & 0.605 \\
\hline
\end{tabular}
\caption{The basic parameters of the networks analyzed empirically.
The number of actors $N_a$, the number of ties $N_t$, the average
degree of actors and ties ($\langle q_a \rangle$ and 
$\langle q_t \rangle$, respectively) and in the one-mode projection
onto actors 
($\langle k \rangle$) and ties ($\langle k_t \rangle$), the 
assortativity coefficient $r$ 
(Eq.~(\ref{eq:pearson_correlation_coefficient})), the clustering coefficient $c$ 
(Eq.~(\ref{eq:clustering_coefficient})), and the average clustering
$\overline{C}$ (Eq.~(\ref{eq:average_clustering})).}
\label{table:parameters}
\end{table}

\subsection{Degree distributions}
\label{sec:degree_distributions}

\begin{figure}[t]
\begin{center}
\includegraphics[width=8cm]{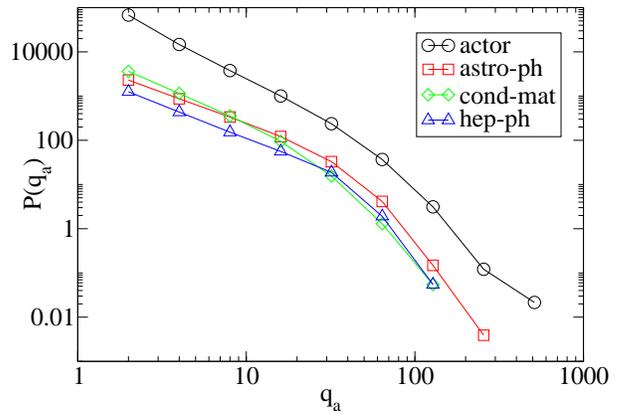}
\caption{The actor degree distributions of the empirical data sets.
All data appear to follow a power-law with exponent -1.6 for small $k$
but there is a noticeable cutoff in each set. }
\label{fig:actordd}
\end{center}
\end{figure}

The degree distributions $P(q_a)$ of the actors in the bipartite
graph form are plotted in 
Fig.~\ref{fig:actordd}. 
Logarithmic binning is used to reduce the effect of statistical
fluctuations.
All the four data sets can roughly be fitted
by a power-law degree distribution $P(q_a)~\sim~q_a^{-\gamma_a}$
with $\gamma_a \approx 1.6$ in the low-$k$ region, but there is 
a pronounced high degree cutoff. This is very similar in all the
cases considered.

\begin{figure}[t]
\begin{center}
\includegraphics[width=8cm]{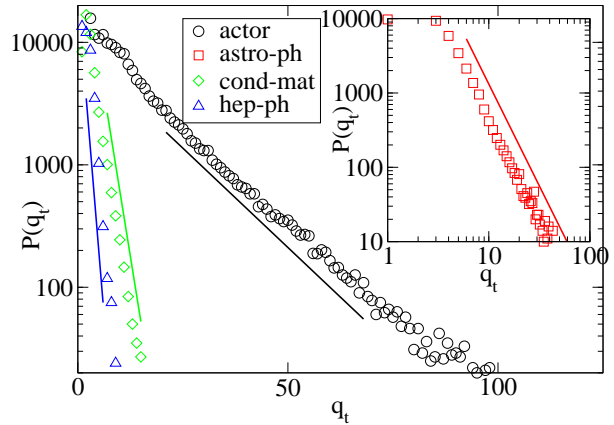}
\caption{The tie degree distributions of the empirical data sets.
The astrophysics collaboration network
seems to exhibit a power-law distribution with the exponent being
approximately -2.7. All the other data sets have an exponential
decay. The solid straight lines are guides to the eye. }
\label{fig:tiedd}
\end{center}
\end{figure}

The degree distributions of the ties $P(q_t)$ are depicted in 
Fig.~\ref{fig:tiedd}. The movie--actor network data and the
cond-mat and hep-ph scientist collaboration networks show an 
exponentially decaying degree distribution 
$P(q_t) \sim e^{-q_t/q_0}$ with $q_0\approx13.4$, $2.0$ and $1.0$ for the
actor, condensed matter and high energy physics data sets, respectively.
The astro-ph
network is an exception since it clearly exhibits a power-law
degree distribution for the ties (the inset of 
Fig.~\ref{fig:tiedd}), i.e. $P(q_t) \sim q_t^{-\gamma_t}$
with $\gamma_t \approx 2.7$. This would seem to point to the direction that
the collaboration patterns in the astrophysics community differ
essentially from those in other disciplines studied here. However, 
as seen below, most of the characteristic quantities of the networks
are unaffected by such a different.

\begin{figure}[t]
\begin{center}
\includegraphics[width=8cm]{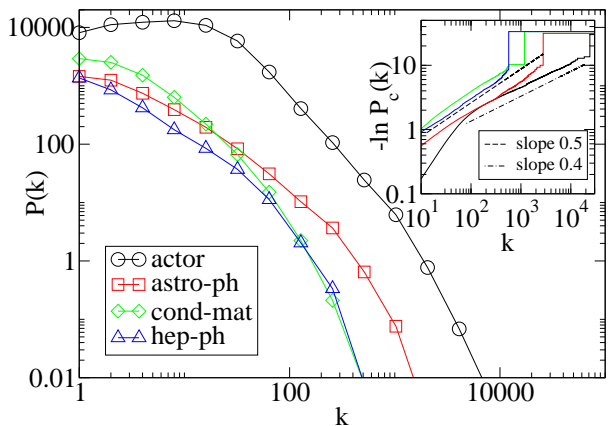}
\caption{The degree distributions in the one-mode projection
onto actors for the different empirical data sets as indicated
by the legend. 
The scientist--article degree distributions
are clearly not scale-free, but more reminiscent of the stretched 
exponential form (Eq.~(\ref{eq:stretched_exponential}))
with $\alpha\approx 0.5$ as indicated in the inset. The movie-actor
network appears to have a power-law scaling regime around $k=100$ but
a more careful examination shows that the tail behavior is also of the 
stretched exponential form, now with $\alpha\approx 0.4$.
The inset uses the same color coding for different data sets as the
main figure.}
\label{fig:ompdd}
\end{center}
\end{figure}

The degree distributions in the one-mode projection onto actors
are shown in Fig.~\ref{fig:ompdd}. 
From the figure we see that the degree distribution of the actor-movie
network
has a lump in the lower-degree region, 
which is also somewhat visible in Fig.~\ref{fig:tiedd}, and
a short power-law region around
$k=100$. However, a careful look reveals that the tail
behavior follows the stretched exponential form 
\begin{equation} \label{eq:stretched_exponential}
P(k) \propto k^{-\alpha}\exp(-\frac{\mu}{1-\alpha}k^{1-\alpha}) \, ,
\end{equation}
where $\mu$ depends on $\alpha$ and satisfies 
$1\le\mu\le2$  \cite{krapivsky:prl85, krapivsky:pre63}. 
For networks with this kind of degree distributions,
the logarithm of the cumulative distribution function
(shown in the inset of Fig.~\ref{fig:ompdd}) is a power-law of the degree
$k$. The inset clearly shows that this is the case here, and that 
in the 
scientist collaboration networks $\alpha \approx 0.5$ and 
in the actor network $\alpha \approx 0.4$. 
The observations made here
about the degree distributions are compatible with previous
studies \cite{ramasco:pre70, newman:pre64r}. 

\subsection{Clustering and correlations}

\begin{figure}[t]
\begin{center}
\includegraphics[width=8cm]{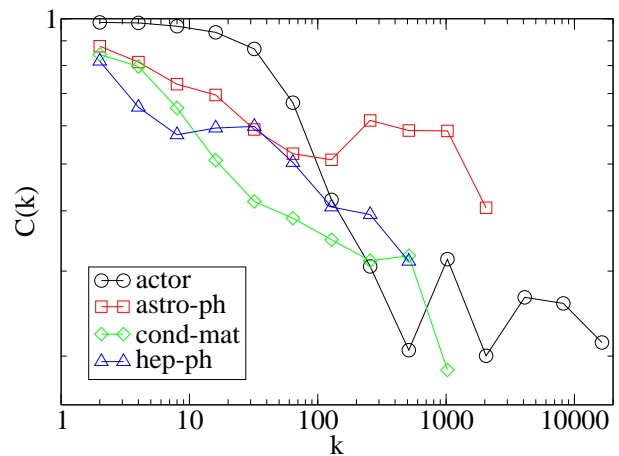}
\caption{The $k$-dependent clusterings of 
Eq.~(\ref{eq:k_clustering}) 
for the empirical data sets as indicated by the legend. 
It is worth noting that for small degrees $k$ the clustering
is huge. The noisiness at
high $k$ ($k>100$ for the scientist collaboration data, $k>1000$ for 
the actor data) comes from the low statistical quality of the data
in these regions (cf.~the degree distributions in Fig.~\ref{fig:ompdd}).}
\label{fig:kclust}
\end{center}
\end{figure}

The degree-dependent clusterings
(naturally, in the one-mode projection)
of Eq.~(\ref{eq:k_clustering}) are
plotted in Fig.~\ref{fig:kclust}, from
which we see that the clustering is substantial
(very close to one) for vertices with small degrees and gets 
lower with an increasing $k$. The low-$k$ behavior is 
expected since the actors with a small $k$ are likely
to be connected only to collaborators sharing a single-tie,
in which case the single-node clustering equals one. 
Furthermore, $C(k)$ is also expected to be monotonically 
decreasing because the more collaborators a node has
the less probable it will be for those to be connected 
with each other.

\begin{figure}[!h]
\begin{center}
\includegraphics[width=8cm]{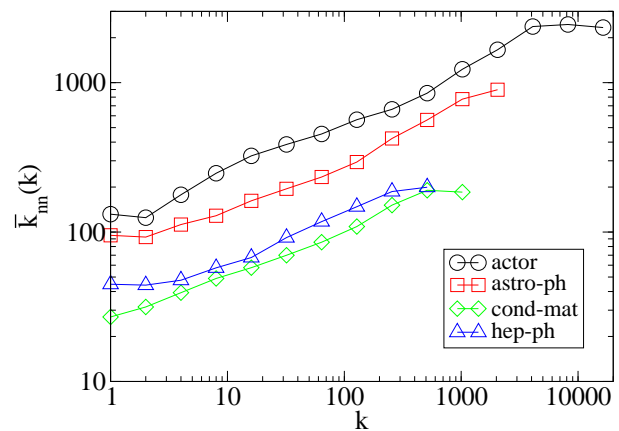}
\caption{The average nearest-neighbour degrees (ANND) 
$\overline{k}_{nn}(k)$ 
as a function of the node
degree $k$ for the empirical data sets. 
For each network, a power-law
with exponent $\beta \approx 0.3$ can be roughly fitted to the data.
This power-law behaviour is the
most important empirical observation made here.
Since the ANND is an increasing function of $k$, there is 
assortative mixing in the networks as typically in social networks. }
\label{fig:knnk}
\end{center}
\end{figure}

The average nearest-neighbour degrees $\langle k_{nn}(k)\rangle$
(Eq.~(\ref{eq:annd_def}))
as a function of node degree $k$ 
are plotted in Fig.~\ref{fig:knnk}. All networks behave similarly with respect
to this quantity; a power-law scaling 
\begin{equation}
\overline{k}_{nn}(k) \sim k^\beta
\end{equation}
with $\beta \approx 0.3$ is observed in each one, with some small deviations.
At high degrees $k$ a cutoff, possibly 
a trace of the finite size of the networks, can be seen in each
data set. 

Since the ANND is an increasing function of $k$, one can conclude that
significant assortative mixing is present in the network, i.e.~the 
degrees of adjacent nodes are positively correlated. 
This can also be seen from the experimentally
measured assortativity coefficient $r$ in Table \ref{table:parameters}.
The differences in the amplitudes of the ANND curves in 
Fig.~\ref{fig:knnk} are explained by the differences in the average
degree $\langle k \rangle$ of the networks.

\subsection{Preferential attachment}
\label{sec:preferential_attachment}

From the empirical data, 
the (one-mode) preferential attachment (PA) rule can also
be measured quite straightforwardly given that the order of 
appearance (or preparation) of the articles or movies can be deduced
from the available data, which is the case here.
The measurement method has been devised by Newman \cite{newman:pre64r}
and goes as follows. 
Denote the degree-dependent preferential attachment rule,
i.e.~the bias to select actors of degree $k$,
by $T_k$. 
Given such a rule, 
the time-dependent probability that 
a node added to the network at time $t$ 
connects to a node of degree $k$ is given by
\begin{equation}
P_k(t) = \frac{T_k n_k(t-1)}{N(t-1)} \, ,
\end{equation}
where $n_k(t-1)$ is the number of nodes with degree $k$ right before the 
addition of the new node and $N(t-1)$ is the total number of nodes in the
graph at the same time. Given these quantities, the preferential 
attachment rule $T_k$ can be measured by making a histogram as a 
function of $k$ to which a new link is added with the weight of 
$N(t-1)/n_k(t-1)$ each time one is created. 

If the attachment is non-preferential, $T_k$ is 
independent of $k$. On the other hand, 
with preferential attachment, $T_k$ is a growing function of the degree $k$
and for instance in the Barab\'asi-Albert model \cite{barabasi:science286} 
$T_k \propto k$ by definition.

\begin{figure*}
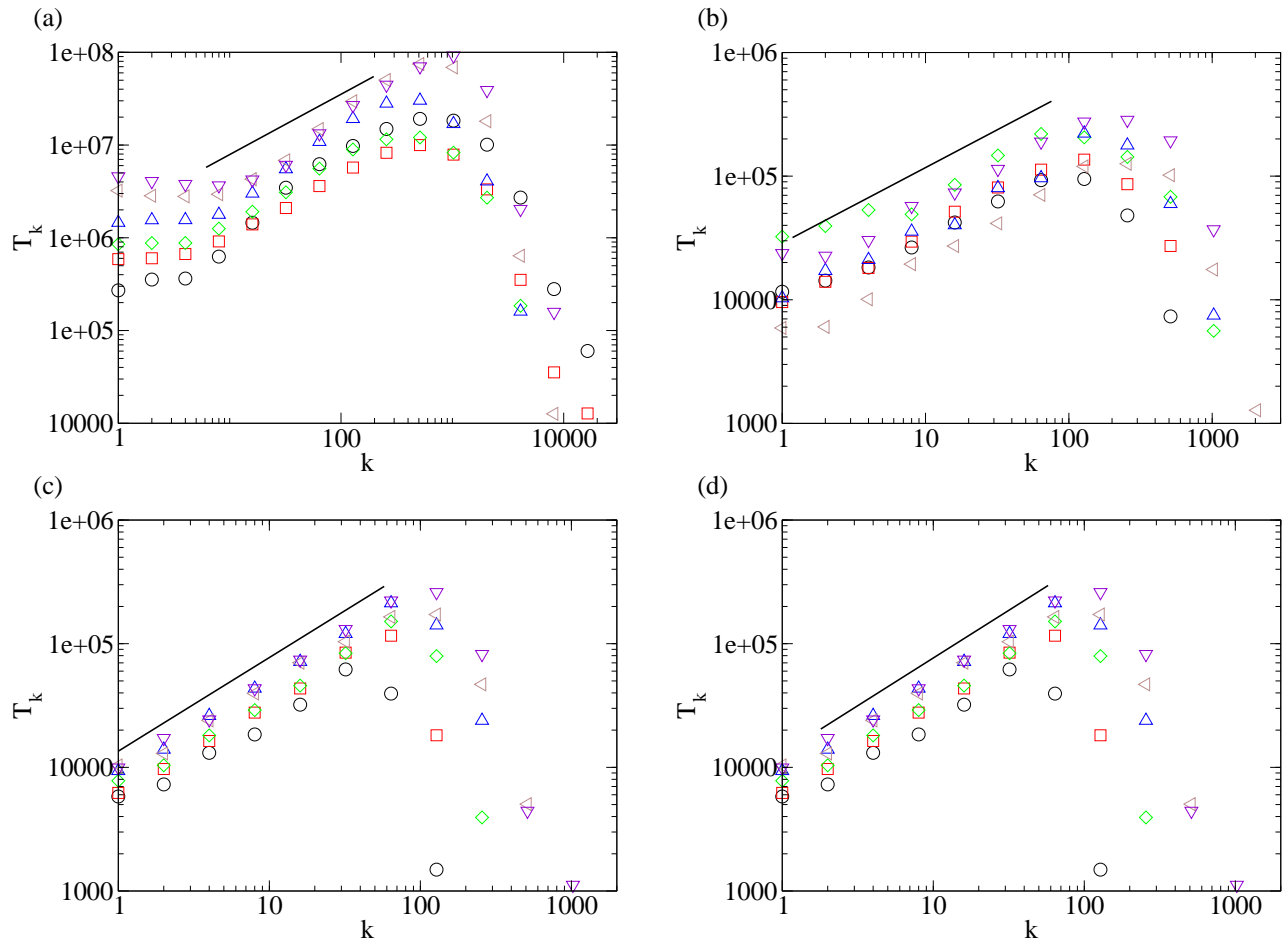

\begin{center}
\includegraphics[width=0.45\textwidth]{figs/parule/actor} \qquad
\includegraphics[width=0.45\textwidth]{figs/parule/astro} \\
\includegraphics[width=0.45\textwidth]{figs/parule/cond} \qquad
\includegraphics[width=0.45\textwidth]{figs/parule/hep} 
\caption{The measured effective preferential attachment (PA) rules for
(a) the actor-movie network in the 
1950s ($\circ$), 1960s ({\color{red}$\square$}), 1970s ({\color{green}$\Diamond$}), 
1980s ({\color{blue}$\vartriangle$}), 1990s ({\color{brown}$\vartriangleleft$}) 
and 2000s ({\color{purple}$\triangledown$}), and for 
(b) the astrophysics, (c) the condensed matter physics and for (d)
the high energy physics scientist--article networks during years
1998 ($\circ$), 1999 ({\color{red}$\square$}), 2000 ({\color{green}$\Diamond$}), 
2001 ({\color{blue}$\vartriangle$}), 2002 ({\color{brown}$\vartriangleleft$}) 
and 2003 ({\color{purple}$\triangledown$}). 
The PA 
rules are well fitted by a power-law $T_k \sim k^\alpha$ with 
a cutoff and they appear to be independent of time. Numerically, we
observe $\alpha \approx 0.65$ for the actor network, 
$\alpha \approx 0.6$ for the astrophysics network and 
$\alpha \approx 0.75$ for the other networks. The solid lines
are guides to the eye.}
\label{fig:parule}
\end{center}
\end{figure*}

The empirically measured PA rules are plotted in Fig.~\ref{fig:parule}.
All of them are well fitted by
power-laws $T_k \sim k^\alpha$ with high-$k$-cutoffs. For the 
actor-movie network the measured value of the exponent 
$\alpha \approx 0.65$, for the 
astrophysics network 
$\alpha \approx 0.6$, and for the other networks $\alpha \approx 0.75$.
Different decades (movies) or years (articles) of accumulation of the data
are shown separately to illustrate that the effective PA rule is 
independent of time. Note that the amplitudes of the plotted curves are
irrelevant, since $T_k$ is a relative probability. 
At low degrees ($k<10$) the behaviour of the actor
data set differs from the other ones. In essence, $T_k$ is approximately
constant in this region. This means that for low degrees the actual
value becomes irrelevant and may perhaps indicate a 
sublinear version of Eq.~(\ref{eq:kplusA}). Note also that we have
observed different exponents $\alpha$ for different data sets but 
the same numerical value of the ANND exponent $\beta$, effectively
ruling out a direct connection between ANND and $T_k$.

The position of the cutoff increases as a function of time, and thus as a
function of the network size. A similar cutoff can be observed
when measuring the PA rule retroactively for a network generated
numerically, so
we conclude that the cutoff is merely a finite-size effect,
which does not need to be taken into account explicitly while building
a simulational model. Similarly, in the team assembly model, the retroactively measured
PA rule is a power-law with a cut-off, but with $\alpha\approx0.4$ and independent of
the simulation parameters. We have not tried to consider the
``$T_k$'' for the tie one-mode projection, though it would
naturally be of some interest.

Measurements of the preferential attachment rule in the arXiv.org 
collaboration networks were also reported in 
Ref.~\cite{newman:pre64r} by Newman. The measurement method used is the same as in
this work. Surprisingly, Newman concludes that the preferential
attachment is linear, which is in striking contrast to the results
obtained here. Since the data sets and the measurement method are 
apparently the same, there remains only one possible explanation. 
Newman considered the arXiv.org 
network as a whole whereas in this work the division into
disciplines is used. Also Barab\'asi \emph{et al.}~\cite{barabasi:physicaa311}
have measured the
PA rule, but for different networks. They discover exponents 0.75 and
0.8 for neuro-science and mathematics scientist--article collaboration
networks, respectively. 

\subsection{The one-mode projection onto ties}
\label{sec:tie_omp_empirics}

\begin{figure}[!h]
\begin{center}
\includegraphics[width=8cm]{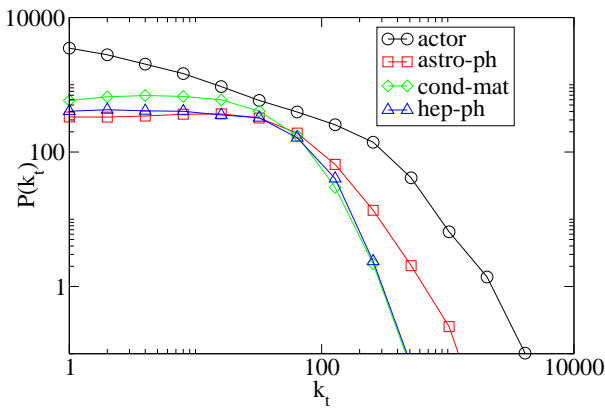}
\caption{The degree distributions in the one-mode projection onto
ties. The scientist-article networks have a region up to 
$k_a \approx 100$ where the probability density is practically a 
constant, followed by a rapid decay. }
\label{fig:tomp_dd_emp}
\end{center}
\end{figure}

\begin{figure}[!h]
\begin{center}
\includegraphics[width=8cm]{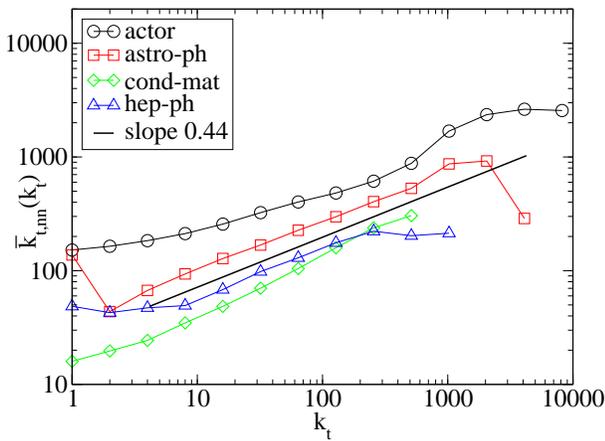}
\caption{The average nearest-neighbour degrees in the one-mode projection
onto ties. A common property of the
scientist collaboration networks is an approximate power-law scaling.}
\label{fig:tomp_knnk_emp}
\end{center}
\end{figure}

The degree distribution and the average nearest neighbour-degree in the
one mode projection onto social ties are shown in 
Figs.~\ref{fig:tomp_dd_emp} and \ref{fig:tomp_knnk_emp} respectively. The 
degree distributions in the scientist collaboration (article) networks are
quite interesting. There is a practically constant region up to 
$k \approx 100$ and a relatively rapid fall at larger degrees. On the other
hand, the movie network appears to behave differently. There is an
approximate power-law with slope around $-0.6$ for small $k$ and 
no region of constant probability can be observed. 

Analogously to the projection onto actors, the average nearest-neighbour
degree approximately scales as a power of the degree $k$ also on this 
projection. The measured exponent is $\beta_t \approx 0.44$. 

\section{Previous models}
\label{sec:earlier_studies}

Earlier studies of collaboration networks are mostly centered around unipartite
networks, with a few exceptions
\cite{ramasco:pre70, borner:pnas101, goldstein:cond-mat0409, guillaume:cond-mat0307,
morris:cond-mat0501, barber:physics0509}. 
Growing unipartite networks are relevant to the present work
since they tell how the degree distribution depends on the growth rule. E.g. if uniform
attachment would be used, the degree distribution becomes exponential
\cite{dorogovtsev:book}, in contrast to a linear preferential attachment rule,
with a power-law degree distribution \cite{barabasi:science286}.
Between these two extremes lies sublinear preferential attachment, i.e.~a network 
growth rule that states that a new node connects to an existing one with
probability proportional to $k^\alpha$ ($0 < \alpha < 1$), where $k$ is the degree
of the existing node. 
This kind of growth leads to a stretched exponential degree distribution
of Eq.~(\ref{eq:stretched_exponential}).

Another family of preferential attachment rules is given by attachment probabilities
$\Pi_k$ of the form
\begin{equation} \label{eq:kplusA}
\Pi_k \propto k + A \, , 
\end{equation}
where $A$ is a parameter also termed additional attractiveness
\cite{dorogovtsev:book}. This leads to scale-free
networks with a tunable, $A$-dependent, degree distribution exponent $\gamma$. 
The networks develop degree correlations such that for $A>0$,
$\overline{k}_{nn}(k) \sim \log(k)$ whereas for $A<0$, a decaying
power-law $\overline{k}_{nn}(k) \sim k^{\beta}$ with $\beta < 0$ is 
recovered \cite{barrat:cond-mat0410}.

The reference model is the bipartite configuration
model \cite{newmanpark:pre68, guillaume:cond-mat0307}. In it, all actors and
ties are created first, assigned degrees from given degree distributions
and linked randomly such that the degrees are fulfilled. It has been proven
mathematically \cite{newman:pre68} that this model always leads to a 
non-negative assortativity coefficient. Despite of this, the correlations
are clearly too weak to explain those in the empirical data \cite{ramasco:pre70}.

Recently Ramasco \emph{et al.} \cite{ramasco:pre70} have introduced a model
of a growing bipartite collaboration network, which is defined as follows. 
At each time step, a new tie with $n$ actors is added to the network. Of these,
$m (<n)$ are new, i.e.~are not currently a part or the network. The rest $n-m$ 
are chosen from the set of pre-existent actors with probability proportional to 
the number of ties $q$ they have already participated in, i.e.~by using linear
preferential attachment. Ramasco \emph{et al.}~arrive at a scale-free degree
distribution $P(k) \sim k^{-\gamma}$ in the one-mode projection onto actors
with
\begin{equation}
\gamma = 2 + \frac{m}{n-m} \, .
\end{equation}

While the model above can be solved analytically, it fails to explain
some features when compared to empirical data \cite{ramasco:pre70}. The
most important deviations are in clustering and correlations. To 
overcome this, Ramasco and co-workers introduced the aging of actors
as an additional property of the model. In their model, the actors
age such that the probability that an actor is alive (i.e.~capable
of participating in new ties) after having participated in $q$ ties
is
\begin{equation} \label{eq:ramasco_aging}
P_{\mathrm{alive}}(q) = 
\left\{ 
\begin{array}{lll}
1 & , & \mathrm{if} \;\;q < q_0 \\
\exp\{-\frac{q-q_0}{\tau}\} & , & \mathrm{if} \;\;q \ge q_0
\end{array}
\right. \, .
\end{equation}
There is a certain survival up to participation in $q_0$
ties and an exponential decay thereafter with a characteristic time $\tau$.

Introducing the aging renders the model analytically unsolvable and
the degree distribution develops a cutoff at large values of the
one-mode degree $k$. A similar cutoff is observed experimentally
and Ramasco and co-workers use its position and steepness for 
determining the values of the aging parameters $q_0$ and $\tau$.
With the aging one is able to make the assortativity coefficient
(Eq.~(\ref{eq:pearson_correlation_coefficient})) positive and
increase the clustering coefficient 
(Eq.~(\ref{eq:clustering_coefficient}))
so that both get closer to empirically observed values.

Guimer\'a \emph{et al.} \cite{guimera:science308} have introduced a model
of team assembly mechanisms that is quite similar to the one discussed above. 
In their model, new teams are formed, and their members are 
selected according to the following rules. At each time step a new team
with $m$ members is created. For each member an incumbent, that is a member that
is already part of the network, is chosen with probability $p$, and a new one
is created with probability $1-p$.  If the previous member was an incumbent,
the next one is selected from its collaborators with probability $q$, otherwise
the selection is performed as above. 

Identifying the teams with the social ties and the team members as the social 
actors, the team assembly model is a version of a growing bipartite network
model. In it, the preferential attachment rule consists of two ingredients:
The first incumbent member of each team and the subsequent ones with probability
$1-q$ are selected with random attachment, whereas the rest are chosen using
linear preferential attachment, which comes here into play implicitly since a
random previous collaborator of a member is chosen \cite{holme:pre65}. 
Guimer\'a \emph{et al.} have found that the degree distribution of the one-mode
projection of the resulting graph mimicks empirically measured distributions
reasonably well \cite{guimera:science308}.  

Somewhat similar studies have also been conducted by 
B\"orner \emph{et al.}~\cite{borner:pnas101}, Goldstein
\emph{et al.}~\cite{goldstein:cond-mat0409} and Morris 
\cite{morris:cond-mat0501}.
A common goal
of these is to introduce realistic models of collaboration
networks. However, they do not pay any attention to 
degree-degree correlations. Similar ideas have also been applied to 
the bipartite network
of research projects funded by the European Union and organizations
participating in them \cite{barber:physics0509}.

\section{Sublinear model}

Motivated by the empirical observations of the previous section, 
the following model of a growing
bipartite collaboration network is proposed. It is also  
related to that of Ramasco \emph{et al.}~\cite{ramasco:pre70}
(see also Sec.~\ref{sec:earlier_studies}), but it is significantly
modified in order for it to be consistent with empirical facts. 

\begin{figure*}[t]
\begin{center}
\includegraphics[width=0.7\textwidth]{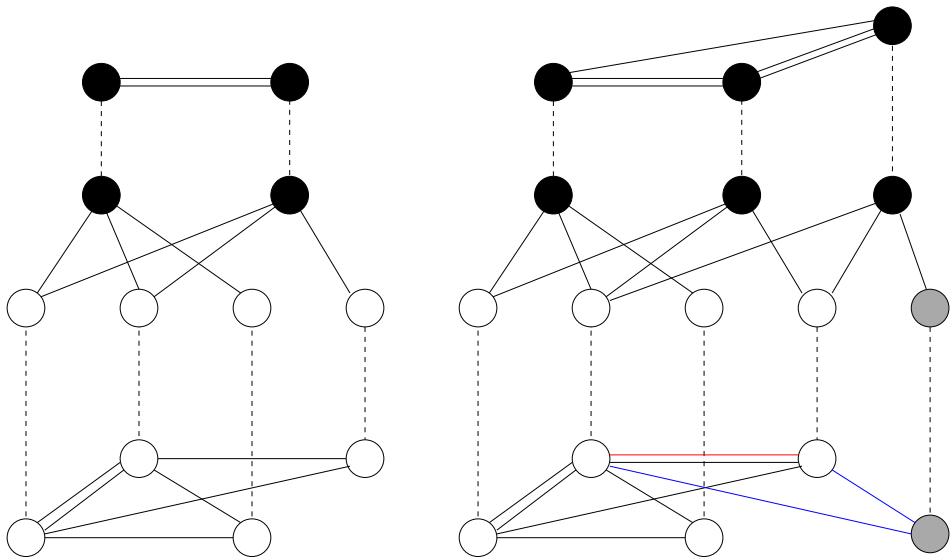}
\caption{An illustration of the one-mode projection and
the addition of a new tie in the model with
sublinear preferential attachment. 
Above the
filled circles denote social ties, the open ones social actors and the
lines the links between them. Below, each social actor is drawn again,
now with links between them in the one-mode projection visible, 
i.e.~two actors are connected if they participate in the same tie.
(a) The network before the addition. For example, the leftmost actor has 
the bipartite degree $q_a=2$ and the one-mode projected degree
$k=4$. Similarly, the leftmost tie has $q_t=3$ and $k_t=2$.
(b) The 
network after the addition. The corresponding one-mode projections
onto actors (ties)
are drawn below (above) the bipartite networks. 
The new tie (the rightmost filled circle) 
introduced one new actor (the grey circle) who acquired two links
to pre-existing actors (shown by blue lines). 
The new tie also caused a new connection between two pre-existing actors
(shown by red line). The new tie connected to both existing ties; once to
the leftmost one (one common actor) and twice to the middle one (two
common actors).
The same growth schematics also
apply to the model of Ramasco \emph{et al.} \cite{ramasco:pre70} and 
to the team assembly model
\cite{guimera:science308}.
}

\label{fig:schematics}
\end{center}
\end{figure*}

The model is as follows (the addition of a new tie is 
illustrated in Fig.~\ref{fig:schematics}).
\begin{itemize}
\item At each time step, a new tie is added to the 
network. The number of actors $n$ of this tie is a random variable
whose distribution is given as an input parameter. Since this
distribution can be measured experimentally, and the measured 
functional form is to be used, there is no fitting involved. 
\item Of these $n$ actors each one has a given probability $p$ to 
be a new one, i.e.~at a fixed $n$ the number of new actors is a 
random variable with binomial distribution. The probability $p$
can be estimated from the data by selecting it such that, on the
average, the total number of actors in the end of the simulation
equals that in the empirical network being mimicked. 
\item The first one of 
the rest $n-m$ of them, with $m$ being the number of new actors,
is chosen from the pool of all pre-existent
actors such that actor $i$ gets chosen with probability
\begin{equation}
p_i = \frac{k_i^\alpha}{\sum_i k_i^\alpha} \, .
\end{equation}
i.e.~with sublinear preferential attachment. The corresponding rule of
the model of Ramasco \emph{et al.} has $\alpha = 1$. 
\item Each one of the rest $n-m-1$ actors is chosen from the set of the
earlier collaborators of the previously chosen actor with probability 
$p_{TF}$ also with sublinear preferential attachment, and as described in
the previous point with probability $1-p_{TF}$. 
\end{itemize}

The most important new ingredient, the sublinear form of the 
PA rule, is justified by the measurements
in Sec.~\ref{sec:preferential_attachment}.
Another quantum of motivation comes from the fact that degree distributions
which are not pure power-laws have been measured 
(see Fig.~\ref{fig:ompdd} in Sec.~\ref{sec:degree_distributions}) and 
such degree distributions have been demonstrated to be caused by
sublinear PA. Still
further motivation is given by the studies of Onody and
de Castro \cite{onody:physicaa336} where sublinear PA was
found to lead to positive degree--degree correlations in terms
of the assortativity coefficient $r$. 

The motivation behind the triad formation (TF) process described
in the last rule is the fact that several models incorporating
this kind of behavior have been found to lead to increased 
clustering, closer to what has been observed in real networks
\cite{holme:pre65}.
A similar ingredient is also present in the team assembly model 
\cite{guimera:science308}, where the parameter
$q$ plays the role of the triad formation probability. 

\section{Numerical results}
\label{sec:numerical_results}

\begin{table}[t]
\begin{tabular}{|l|r|r|r|r|}
\hline
network & empirical & sublinear & sublinear & team assembly\\
& & $p_{TF}=0.0$ & $p_{TF}=0.9$ & \\
\hline \hline
$N_a$ & 28 526 & 25 497 & 24 650 & 24 477\\
$N_t$ & 49 330 & 49 000 & 49 000 & 49 000\\
$\langle q_a \rangle$ & 5.08 & 6.65 & 6.92 & 6.95 \\
$\langle q_t \rangle$ & 2.94 & 3.46 & 3.48 & 3.47 \\
$\langle k \rangle$ & 16.0 & 13.8 & 14.1 & 13.9 \\
$\langle k_t \rangle$ & 56.3 & 36.1 & 107.6 & 33.2 \\
$r$ & 0.250 & 0.13 & 0.15 & 0.26 \\
$c$ & 0.370 & 0.10 & 0.18 & 0.32 \\
$\overline{C}$ & 0.674 & 0.51 & 0.66 & 0.65 \\
\hline
\end{tabular}
\caption{The basic parameters of the simulated networks compared
with the empirical condensed matter collaboration network.}
\label{table:parameters_simulation}
\end{table}

In the simulations reported in this section, the number of ties in a simulated
network is always $N_t = 49000$, and the fraction of new social actors in a
given tie is $p_{new} = 0.202$. The same parameters are also used for simulations
of the team assembly model, i.e.~
$p = 1-p_{new}$, and the simulation
is run until $N_t$ teams are created.
This selection of the parameters comes directly
from the empirical measures of the condensed matter collaboration network. In the simulations
of the team assembly model, the probability to select a previous collaborator
of an incumbent is $q = p_{TF} = 0.9$ unless otherwise mentioned. This choice leads
to the correct order of magnitude in the clustering of the resulting graph, as
will be seen below. In both models, the number $n$ of actors in a tie a drawn from 
the same probability distribution that corresponds to the empirically measured
one (see Fig.~\ref{fig:tiedd}). In addition, the aging mechanism is omitted in both
models, i.e.~the parameter $\tau$ of the team assembly model and the parameter $q_0$
of the model of Ramasco \emph{ et al.}~are set to infinity. This appears to be 
justified since the ubiquitous characteristics of the ANND and the $k$-dependent
clustering $C(k)$ do not show experimental dependence on the network age. Indeed, 
these are similar for both the movie network (in which aging surely has taken place)
and the physics collaboration networks, in which the data collection is for a short
interval and aging plays only a little role if at all. All the simulation results
are from a single simulation run, i.e.~from one network. To check the
validity of this approach, we ran several simulations with the same parameters. The
simulations are 
practically indistinguishable from each other, and thus we conclude that
this is justified.

The basic parameters of the simulated networks, compared with the empirical
condensed matter data, are shown in Table \ref{table:parameters_simulation}.
The parameters that are either input to the models or straightforwardly
depend on those, such as the number of nodes of different kinds, the 
average degrees in the bipartite network and in the one-mode projection
onto actors, are, naturally, reproduced quite well. On the other hand, 
already the average degree $\langle k_t \rangle$ in the one-mode projection
onto ties shows discrepancies between the simulations and the data. It 
appears that $p_{TF}$ could be used in the sublinear model to tune this 
value, but in this work the role of the tuning target is played by
the average clustering $\overline{C}$. Regarding the clustering and 
correlations, the best numerical fit is given by the team assembly model. 

\begin{figure}[t]
\begin{center}
\includegraphics[width=8cm]{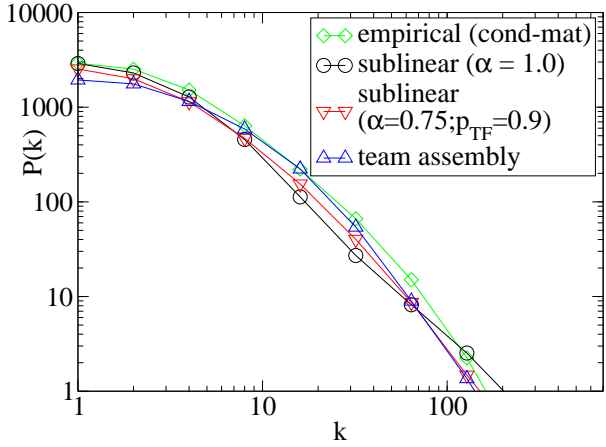}
\caption{The degree distribution in the one-mode projection onto actors
in the condensed matter collaboration network and in simulations of 
the sublinear model with 
different values of the PA exponent $\alpha$ and of the team assembly model
with $p_{TF}$ = 1.0. The curve for $\alpha=1.0$
leads to power-law degree distribution as expected \cite{ramasco:pre70}, 
whereas that with $\alpha=0.75$ and the one of the team assembly model
is clearly a lot closer to the empirical values. In the sublinear model, 
the behavior is the same even for $p_{TF}$=0 (not shown).}
\label{fig:numerical_dd}
\end{center}
\end{figure}

The degree distributions in the one-mode projection onto actors 
are plotted in Fig.~\ref{fig:numerical_dd} for the empirically measured
condensed matter collaboration network and for simulations of the 
sublinear model with
different values of $\alpha$ together with a simulation of the team assembly model. It 
is seen that the simulation of the sublinear model 
(with or without triad formation)
with $\alpha$ equal to the experimentally
measured effective one (Fig.~\ref{fig:parule}) mimicks the empirical
degree distribution significantly better than the one using the linear
PA rule.
The latter leads to a scale-free degree
distribution as predicted \cite{ramasco:pre70}. Also the team assembly model
is capable of reproducing the empirical degree distribution reasonably well.
In the rest of this paper, the value $\alpha=0.75$ is used unless otherwise 
mentioned. Note that the comparison of the other networks studied empirically
in this work to corresponding simulation yields the same behavior. 

\begin{figure}[t]
\begin{center}
\includegraphics[width=8cm]{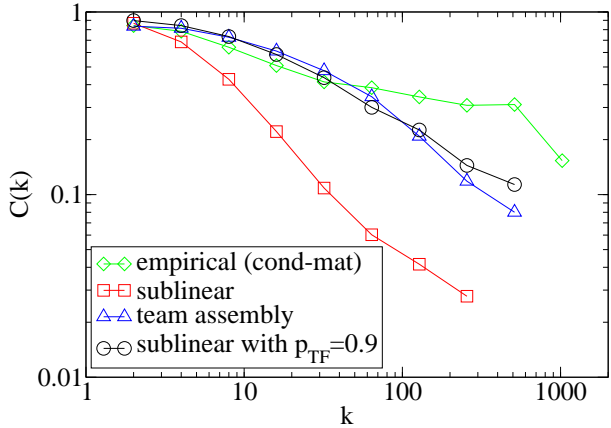}
\caption{The $k$-dependent clustering (Eq.(\ref{eq:k_clustering})) compared to empirical measurements
for both the sublinear model ($\alpha=0.75$) and for the team assembly model. 
It is seen that the sublinear model at $p_{TF}=0$ cannot reproduce the clustering
as expected whereas both the team assembly model and the sublinear model at $p_{TF}=0.9$
do somewhat better. However, not one of the models agrees fully with the empirical data.}
\label{fig:numerical_kclust}
\end{center}
\end{figure}

The degree-dependent clustering $C(k)$ of Eq.~(\ref{eq:k_clustering}) 
in the one-mode projection
is plotted in Fig.~\ref{fig:numerical_kclust} for the condensed matter 
collaboration network and for simulations both of the sublinear model 
and the team assembly model. From
the figure, it can be seen that the sublinear model without triad 
formation
differs notably from the empirical data whereas the sublinear model with
a high probability for triad formation and the team assembly model give
a correct order of magnitude for the overall clustering 
(see also Table \ref{table:parameters_simulation})
but the form of the $C(k)$ curve differs from the empirical one. 
In this respect,
the sublinear model does slightly better that the team assembly model.

\begin{figure}[t]
\begin{center}
\includegraphics[width=8cm]{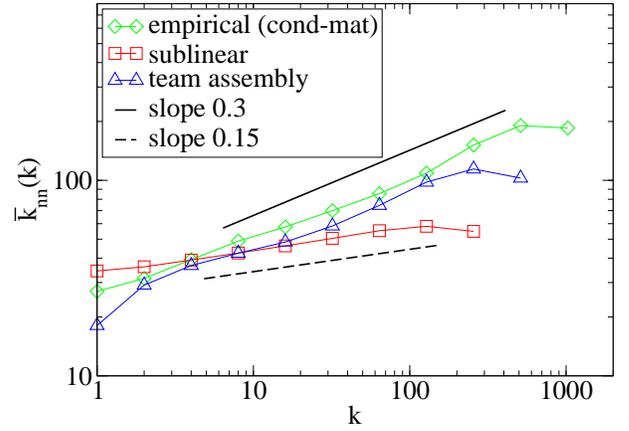}
\caption{The average nearest-neighbour degree (ANND) in simulations of the sublinear
model ($\alpha=0.75$)
and the team assembly model compared to the empirical measurements. The results
from the team assembly model are in reasonable
agreement with the real data, whereas those
from the sublinear one roughly scale as a power of $k$ but with a different exponent.
The solid and dashed lines are guides to the eye.}
\label{fig:numerical_knnk_comparison}
\end{center}
\end{figure}

\begin{figure}[t]
\begin{center}
\includegraphics[width=8cm]{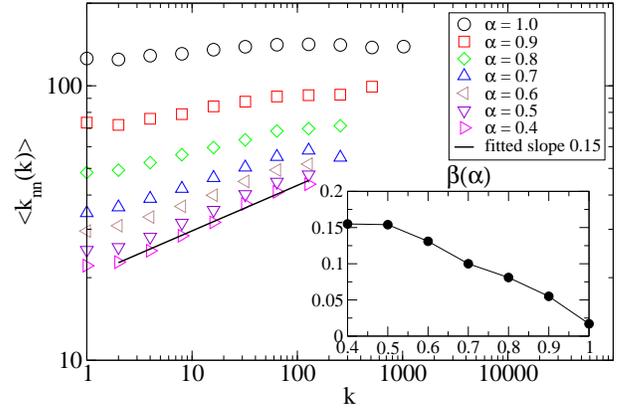}
\caption{The average nearest-neighbour degree (ANND) 
in simulations using different values of the preferential attachment
exponent $\alpha$. For all $\alpha$ the ANND scales as a power of 
the node degree $k$, $\overline{k}_{nn}(k) \sim k^{\beta(\alpha)}$
so that for $\alpha = 1$, $\beta=0$, i.e.~no degree-degree correlations
are observed, and for $\alpha < 1$, $\beta$ grows as $\alpha$ decreases
leading to positive correlations. The inset shows the dependence of $\beta$
on $\alpha$.}
\label{fig:numerical_knnk}
\end{center}
\end{figure}

The average nearest-neighbour degree $\overline{k}_{nn}(k)$ (ANND) is
plotted in Fig.~\ref{fig:numerical_knnk_comparison} for the condensed 
matter empirical data set and for simulations of both models. The figure
shows that the team assembly model reproduces the correlation
structure of the empirical network reasonably well in the 
intermediate-$k$ regime: both appear to roughly scale
as $\overline{k}_{nn}(k) \sim k^\beta$
with $\beta=0.3$ and approximately agree in the amplitude. However, 
the simulation differs from the data at both low and high $k$-values.
On the other hand,
the simulations of the sublinear model show a similar scaling but with
a different, smaller exponent ($\beta \approx 0.15$).

To study the effect of the exponent $\alpha$ on the scaling of the ANND
in the sublinear model, it is 
depicted in Fig.~\ref{fig:numerical_knnk} as a function of $\alpha$.
For $\alpha=1$, we see that there are 
no degree-degree correlations at all, corresponding to the
model of Ramasco and co-workers without aging. For lower values of 
$\alpha$, positive correlations are present and the ANND scales as a 
power-law of the vertex degree $k$ as above.
The value of $\beta$ depends
continously on $\alpha$ as seen in the inset of Fig.~\ref{fig:numerical_knnk}.
However, the numerical value of the exponent 
$\beta$ is notably lower in the relevant region
$0.6 < \alpha < 0.8$ than the experimentally observed one.
Thus, the overall correlations in this case are not as strong as in
the empirical data. This conclusion can also arrived at considering
the values of the assortativity coefficient $r$ in 
Table \ref{table:parameters_simulation}.

\begin{figure}[t]
\begin{center}
\includegraphics[width=8cm]{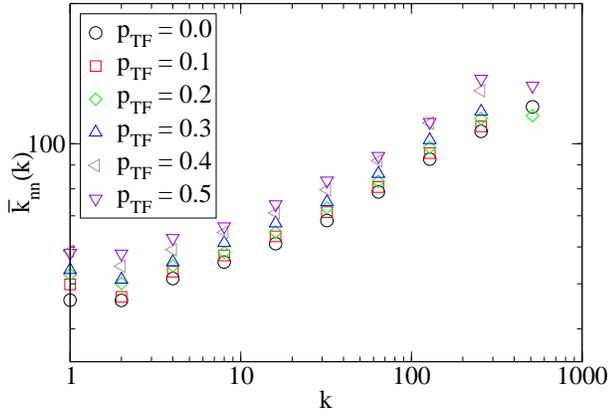}
\caption{The scaling of the ANND for different values of the 
triad formation (TF) probability in simulations of the sublinear model. 
It is seen that introducing a TF mechanism does not change the scaling.
In addition to the results in the figure ($\alpha=0.7$) we have checked
this with several other values of $\alpha$, too, with the same results.
The same also applies to the team assembly model (results not shown).}
\label{fig:numerical_holmekim}
\end{center}
\end{figure}

To see how to the triad formation affects the scaling of the ANND,
it is plotted for several values of $p_{TF}$ in
Fig.~\ref{fig:numerical_holmekim}
for the sublinear model.
It is clearly seen from the figures that the triad formation
process has no effect at all. We have also performed a corresponding series
of simulations with several different values of $\alpha$. In all cases, the
conclusion remains the same. 
Simulations of the team assembly model
also revealed the same behaviour. 
Thus, we conclude that
this kind of process can not be held responsible for the observed correlations.
Again, comparing the assortativity coefficient $r$ in Table 
\ref{table:parameters_simulation} for the sublinear model with and without
triad formation supports this conclusion.

\begin{figure}[t]
\begin{center}
\includegraphics[width=8cm]{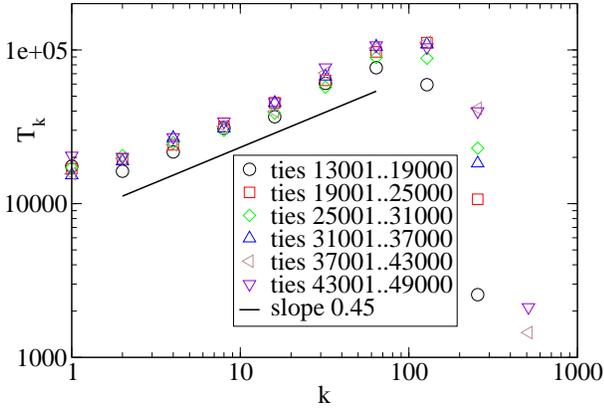}
\caption{The retroactively measured preferential attachment rule for a simulation
of the team assembly model. The rule is approximately a time-independent sublinear
power-law with $\alpha\approx 0.45$ and with a cut-off. The solid line is a guide 
to the eye.}
\label{fig:teamass_pa}
\end{center}
\end{figure}

The retroactively measured preferential attachment rule for the 
team assembly model is plotted in Fig.~\ref{fig:teamass_pa}. It is clearly seen
that the rule is, again, a sublinear power-law with $\alpha \approx 0.4$ and with 
a cut-off that is very similar to the empirical ones (cf.~Fig.~\ref{fig:parule}) and to those
in the simulations of the sublinear model (not shown). This is
surprising since, at the first sight, one could anticipate that the combination
of attachment rules in the team assembly model leads to a compound rule of the
form $k+A$. However, the bipartiteness comes into play, and this phenomenon 
shows that it can indeed affect essentially the network structure. It can also be
seen that the effective attachment rule is time-independent,
as is true concerning the empirical data.

\begin{figure}[t]
\begin{center}
\includegraphics[width=8cm]{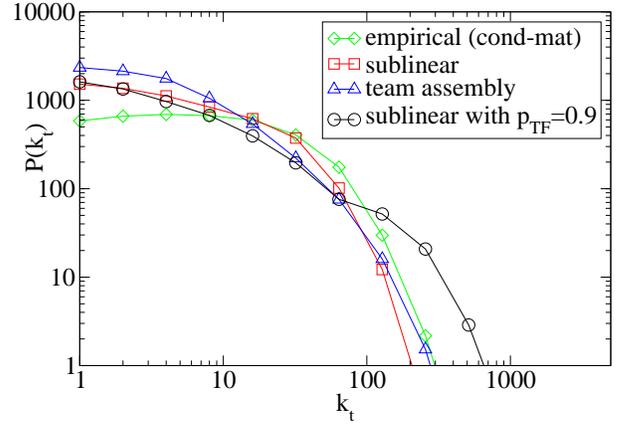}
\caption{The degree distribution in the one-mode projection onto ties
in the empirical condensed matter data set, compared to simulations.
A considerable difference exists between the simulations and the data.}
\label{fig:tieomp_pk}
\end{center}
\end{figure}

\begin{figure}[t]
\begin{center}
\includegraphics[width=8cm]{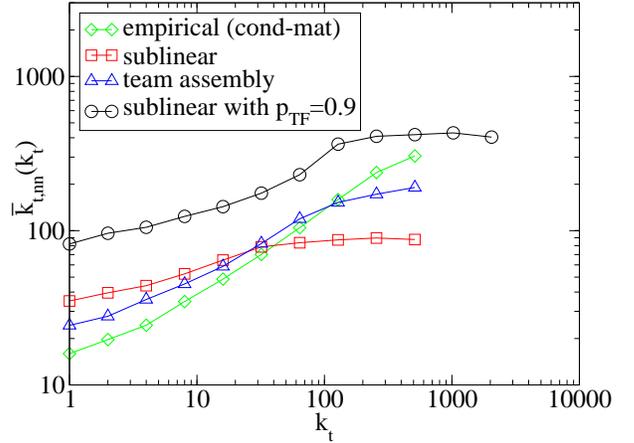}
\caption{The average nearest-neighbour degree in the one-mode projection
onto ties in the empirical condensed matter data set
and in the simulations. These are unable to reproduce the empirically
observed power-law scaling. Quantitatively, the team assembly
model mimicks the behavior of the data better than the sublinear model.}
\label{fig:tieomp_annd}
\end{center}
\end{figure}

Next we compare the models with the one-mode projection onto social ties
instead of social actors. These
are shown in Figs.~\ref{fig:tieomp_pk} and \ref{fig:tieomp_annd} for the degree
distribution and the average nearest-neighbour degree, respectively. 
From Fig.~\ref{fig:tieomp_pk} one can see that the characteristic plateau
of the scientist collaboration networks is not captured by any of the 
simulations. Similar conclusions can be made from Fig.~\ref{fig:tieomp_annd}:
not one of the models lead to the power-law scaling that we have
empirically observed. There are also considerable differences in the 
overall magnitudes of the quantities, in which respect
the team assembly model behaves best. 

\section{Summary and discussion}

In this paper, we have analyzed several bipartite collaboration networks 
empirically. The static and dynamic structure thereof is one of the
conceptually simplest examples of complex networks, where effective
statistical laws seem to exist. The quantitative description of such phenomena
by models becomes then the (elusive) goal. These systems are 
very clear-cut in that the old graph structure is static - old vertices
and edges are not removed - and that the growth events are easy
to quantify by various measures and to follow, from data.

Concerning the correlation structure of collaboration
networks, the most 
important empirical observation is that the
average nearest-neighbour degree (ANND) 
in the one-mode projection onto social actors
scales as a power of the node degree
as $\overline{k}_{nn}(k) \sim k^\beta$ with $\beta \approx 0.3$. 
Similar scaling is also present in the projection onto social ties, 
i.e.~articles or movies. The clustering of the one-mode network(s)
is considerable. The effective actor-projection
preferential attachment (PA) rule appears to be a sublinear
power-law, and independent of time. 

We have also introduced a model, which is built on top of this observation, in an 
attempt to explain the form of the observed properties of the networks. The 
empirically observed sublinearity of the PA rule has thus been included
in a numerical model.
In this case, the ANND indeed scales as a power of $k$,
but the numerical values of the exponents do not match. In any case,
the model is capable of demonstrating that the form of the PA rule can essentially
affect the correlation structure. 

Another model of team assembly mechanisms \cite{guimera:science308} has also
been simulated. The ANND seems to fit the (actor) empirical observations
reasonably well: both roughly scale as $k^\beta$ with $\beta \approx 0.3$. 
A common feature of both models is that they reproduce the degree distribution
in the one-mode projection rather well (see Fig.~\ref{fig:numerical_dd}). 
In the case of the sublinear model, using the correct, 
empirically measured, value
of the exponent $\alpha$ is necessary for this result. 
On the other hand, team assembly model fails to reproduce the
empiricál attachment rule..
The sublinear model without any triad formation fails to reproduce 
the form of the $k$-dependent clustering, whereas the team assembly model
and the sublinear model with considerable probability for triad formation
lead to correct order of magnitude of the average clustering $\overline{C}$,
seen also in the overall magnitude of $C(k)$. However, the models do not
explain its functional form. Note that a triad formation
process does not change the correlations in the models studied here.

Considering the one-mode projection onto social ties instead of actors reveals
the inadequacy of the both models. The empirically measured degree 
distribution and the average nearest-neighbour degree both differ from
their simulated counterparts. In effect, we have observed that even though
various can reproduce some properties of the projections onto actors, 
they lack explanatory power when it comes to considering the networks with 
their full bipartite structure intact. Perhaps one should consider tie-based
growth rules instead of actor-based ones.

Summarizing, there is a clear need for a more complex bipartite
growth model that accounts for both the clustering and correlations
of actors (authors) and ties (articles). Since the bipartite structure
changes by events in which one tie is introduced together with several
actors, this means that the old actors' effective
choice must follow from a rule
that measures the correlation structure in more detail. One candidate
would be to use $k$-connected cliques in analogy to recent observations
of the role of such in network superstructure \cite{palla:nature435}. 
This would allow for various ways of measuring the
joint strength of interaction between old actors. 
Furthermore, using the recently introduced concept of social inertia 
\cite{ramasco:physics0509} might be of use in this respect, by
establishing a quantitative time-dependent measure.
It is also clear
that there are substructures within subfields. These point towards
the idea that the actors and ties have ``hidden variables'' that
should be taken into account. One practical prospect would be to
use e.g.~the PACS indices to classify ties (articles) and actors/authors,
and investigate the role of the both above ideas. 
Note that in all the cases here the ``invisible college'' or giant 
component of the
one-mode projection onto actors includes really almost all of the actors
and is thus trivial. 
It is an open question how to define and measure the ``success'' of an
actor given this, and the performance of current models - simple
membership is not enough. Again, possibly progress could be made
by the use of weighted networks.

Even though several sources 
of positive degree-degree correlations have been demonstrated here, 
there are still open questions related to these. Most importantly, the
reason or origin of the specific form of the correlations remains unknown.
Perhaps one needs to define more informative quantities for measuring
the structure of the original bipartite network.
Studies on how the form of the (one-mode) PA rule depends on the underlying 
elementary social phenomena offer interesting avenues for future work.

{\bf Acknowledgments.} 
We thank Sergey Dorogovtsev for numerous stimulating and useful
discussions, and for a critical reading of an early version of this manuscript. 
This work was supported by the Academy of Finland, Center of Excellence 
program.

\end{document}